\documentclass{article}

\usepackage[dvipsnames]{xcolor} 
\usepackage{colortbl} 
\usepackage{amsmath, amsfonts, amssymb, graphicx, wrapfig, arydshln, amsthm, multirow, booktabs, caption, subcaption, easybmat, bigdelim, listings, afterpage, fancybox, tikz, lscape, mathtools, lmodern, rotating, adjustbox, enumitem, blkarray, cases, listings, longtable}
\usepackage{enumitem} 
\usepackage[toc,page]{appendix} 
\usepackage[english]{babel}
\usepackage{mathpazo} 
\usepackage[super,semicolon]{natbib} 
\usepackage[colorlinks=true,allcolors=blue]{hyperref} 
\usepackage[left=2cm,right=2cm,top=2cm,bottom=2cm]{geometry}
\usepackage{authblk} 

\usetikzlibrary{arrows,positioning,shapes.geometric}

\makeatletter
\def\hlinewd#1{%
  \noalign{\ifnum0=`}\fi\hrule \@height #1 \futurelet
   \reserved@a\@xhline}
\makeatother

\newcommand{\E}{\mathbb{E}} 
\newcommand\bs[1]{\boldsymbol{#1}} 
\newcommand{\constone}{\text{const}_1} 
\newcommand{\consttwo}{\text{const}_2} 
\newcommand{\constthree}{\text{const}_3} 
\newcommand{\logit}{\text{logit}} 

\title{Using smoothing splines to resolve the curvature identifiability problem in age-period-cohort models with unequal intervals.}

\author[1]{Connor Gascoigne (c.gascoigne@bath.ac.uk)}
\author[1]{Theresa Smith}
\affil[1]{Department of Mathematical Sciences, University of Bath, Bath, BA2 7AY, UK}

\date{} 

\begin{document}

\maketitle

\begin{abstract}
Age-period-cohort (APC) models are frequently used in a variety of health and demographic related outcomes. Fitting and interpreting APC models to data in equal intervals (equal age and period widths) is non-trivial due to the structural link between the three temporal effects (given two, the third can always be found) causing the well-known identification problem. The usual method for resolving the structural link identification problem is to base a model off identifiable quantities. It is common to find health and demographic data in unequal intervals, this encounters further identification problems on top of the structural link. We highlight the new issues by showing quantities that were identifiable for equal intervals are no longer identifiable for unequal data. Furthermore, through extensive simulation studies, we show how previous methods for unequal APC models are not always appropriate due to their sensitivity to the choice of functions used to approximate the true temporal functions. We propose a new method for modelling unequal APC data using penalised smoothing splines. Our proposal effectively resolves any additional issues that arise and is robust to the choice of the approximating function. To demonstrate the effectiveness of our proposal, we conclude with an application to UK all-cause mortality data from the Human mortality database (HMD).

\textit{keywords:} Age-period-cohort models, identifiability, penalised smoothing splines, unequal intervals
\end{abstract}

\section{Introduction}
\label{Section: Introduction}

Age-period-cohort (APC) models are used to interpret the effect of the most influential temporal trends on incidence and mortality rates for a multitude of diseases. Age effects are a measure of attrition on one’s body as they get older, period (time of the event) effects reflect short term exposures (e.g., new treatments) and a (commonly birth) cohort effect is a long-term exposure (e.g., smoking views). We use obesity as an example of how all three temporal effects relate to a major health concern. Obesity is a measure of an individuals body mass index (BMI) with those greater than or equal to 30 being classified as obese. The most recent health survey from the UKs national Health Service (NHS) found 68\% and 60\% of adult men and women are classified as obese, respectfully. \citep{NHS2020} Weight of an individual increases with age and in recent years there has been an increasing trend of obesity. These together make for a cohort effect where more those born in more recent cohorts have an increase risk to being obese and getting there at an earlier age.

APC models are affected by a identifiability problem due to the linear dependence between the three temporal terms. For example, a birth cohort can always be found by subtracting the age of the individual from the year the response was taken. The result is rather than three independent time trends, we only have two whose linear trends are impossible to disentangle from one another without the use of additional information. We will call this problem the ``structural link identification problem'' (or structural link for short). 

Commonly, APC models are considered when data comes tabulated in equal widths (equal intervals). Appropriate solutions to the structural link are based on re-parameterising the APC modeling into estimable quantities. When time is considered discrete (most common), each temporal term is modelled as a factor with levels for each time interval; Holford pioneered a solution based on estimable curvatures \cite{holford1983estimation} (terms that are orthogonal a linear term) for each temporal effect whilst Kuang, Nielson and Nielson based a solution on estimable second differences \citep{kuang2008a} (a discrete version of the second derivative). When time is considered continuous, the temporal terms are modelled by approximate smooth functions. Carstensen defined a set of estimable quantities, similar to Holfords curvatures, to fit APC models. \citep{Carstensen2007age} Smith and Wakefield offer a comprehensive review on APC models for data aggregated in equal intervals. \citep{smith2016review}

\begin{figure}[!h]
	\caption{Period curvature estimate from fitting an APC model re-parameterised into linear terms and their orthogonal curvatures to simulated unequal interval APC data.}
	\label{Fig: period_cylcing_intro}
	\centering
	{\includegraphics[width=.75\textwidth]
	{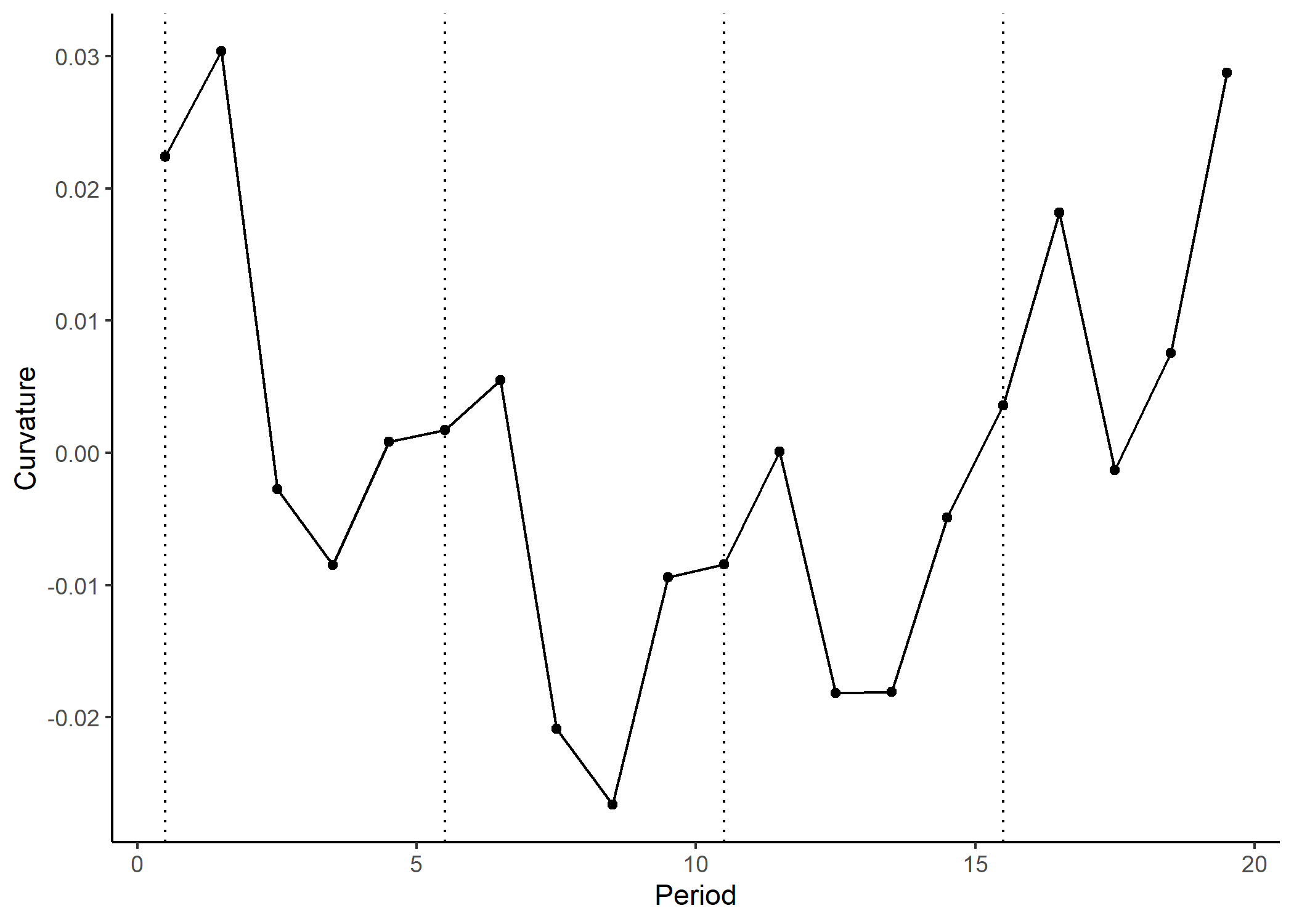}}
\end{figure}

Less commonly, APC models are considered when data comes tabulated in unequal intervals, this contrasts the fact most providers of health and demographic data frequently release data tabulated in unequal intervals. For example, the UK's office of national statistics (ONS) releases all-cause mortality data in single-year age and period \citep{ONS2020yearly} and, for a finer understanding of seasonality, weekly periods and five-year age groups. \citep{ONS2021weekly_period} APC models fit to unequal data are less common as the model fitting process induces more identification issues (on top of the structural link) that are displayed by a cyclic pattern in the previously estimable functions. \citep{holford2006approaches} Figure \ref{Fig: period_cylcing_intro} shows the period curvature estimates when a factor model is fit to simulated unequal data. Note the cyclic pattern, this is could be due to the underlying phenomena of interest being modelled. However, later in the paper it is shown the cyclic pattern is caused by the the further identification problems present when modelling unequal data. 

As Figure \ref{Fig: period_cylcing_intro} alludes to, factor-based approaches based on estimable quantities that worked for data in equal intervals are no longer appropriate when data comes in unequal intervals. A proposed method to model APC data in unequal intervals is to to model the temporal terms with approximate smooth functions, \cite{holford2006approaches, Carstensen2007age} but these do not consider all the issues present fully. Another approach is to collapsing unequal intervals into equal intervals but in many cases this causes a large amount of information lost which decreases the reliability of the results.

The purpose of this paper is to propose a method to modelling APC data that comes in unequal intervals. The method we propose addresses all identification problems present for unequal data, maintains clarity on what is and is not estimable, has a clear interpretation, and is robust to the choice of function used to approximate each temporal term. We propose approximating each temporal term as a continuous function, re-parameterise each into a linear and orthogonal curvature and when modelling the curvatures, include a penalty on the second derivative (a measure of ``wiggliness'') of the estimate. Using continuous functions to model the temporal terms and/or re-parameterising the temporal terms into estimable quantities is not new for APC modelling, but including a penalty term on the curvature is. Via extensive simulation studies, we show the penalty term applied to the estimates of curvature ensure the results from modelling unequal data are robust and do not suffer from any additional issues, unlike the results from the use of continuous functions alone (without a penalty terms) and the factor model.

The remainder of the article is organised as follows. In Section 2, we review the identification problem for data aggregated in equal intervals and introduce our new re-parameterisation scheme. Section 3 is a comprehensive simulation study for the case when data comes in equal intervals. Section 4, we review the curvature identification problem that arises from unequal intervals and show through theoretical and simulation results how the proposed method relieves this added identification problem. Finally, we conclude with an application to all-cause mortality data in the UK in Section 5 and a conclusion in Section 6. 

\section{Method}
\label{Section: Method}

\subsection{Identification Problems}

We begin by discussing an APC model for data equally aggregated, referred to as `equal intervals'. There are two types of identification problems in this model. The first is well-known and due to including an intercept along with more than one smooth function (or factor) in a model. The second and more serious is due to the structural link. The structural link occurs since given any two of age, period or cohort, the third can always calculated. Commonly, birth cohort is found by taking the difference between year of event and age, $c = p - M \times a$ where $M$ is the ratio of age interval to period interval. For equal intervals $M=1$, this simplifies to $c=p-a$.

Table \ref{Tab: M=1 ap data table} shows how cohort index varies when age and period are aggregated into equal intervals. With age increasing from bottom to top and period left to right, a cohorts progression can be traced on the bottom left to top right diagonal. The earliest cohort is top left (oldest age with the first year) and the most recent cohort is bottom left (youngest age with most recent year).
\begin{table}[!h]
	\caption{Cohort indexing for age-period data table where age is grouped $M=1$ times larger than period. The cohort index is defined using $c=M\times\left(A-a\right)+p$ where $A=8$ to fix the first cohort to be 1.}
	\label{Tab: M=1 ap data table}
	\centering
	\begin{tabular}{ccccccccc}
		\hlinewd{2pt} 
		8 & 1 & 2 & 3 & 4 & 5 & 6 & 7 & 8 \\
		7 & 2 & 3 & 4 & 5 & 6 & 7 & 8 & 9 \\ 
		6 & 3 & 4 & 5 & 6 & 7 & 8 & 9 & 10 \\ 
		5 & 4 & 5 & 6 & 7 & 8 & 9 & 10 & 11 \\
		4 & 5 & 6 & 7 & 8 & 9 & 10 & 11 & 12 \\
		3 & 6 & 7 & 8 & 9 & 10 & 11 & 12 & 13 \\
		2 & 7 & 8 & 9 & 10 & 11 & 12 & 13 & 12 \\   
		1 & 8 & 9 & 10 & 11 & 12 & 13 & 14 & 15 \\ 
		\hlinewd{1.5pt}
		\textbf{Age} & 1 & 2 & 3 & 4 & 5 & 6 & 7 & 8 \\
		\cline{2-9}
		& \multicolumn{8}{c}{\textbf{Period}} \\
		\hlinewd{2pt}
	\end{tabular}
\end{table}

Let $y_{ap}$ be response from age group $a$ and period group $p$ where $a = 1, 2, \dots, A$ and $p = 1, 2, \dots, P$. A cohort index is not explicitly defined due to the structural link but is calculated $c = 1, 2, \dots, C = M\times\left(A - a\right)+p$. A continuous APC model is
\begin{equation}
	\label{Eq: Over parameterised APC model}
	g\left(\mu_{ap}\right) = f_A\left(a\right) + f_P\left(p\right) + f_C\left(c\right)
\end{equation}
where $g\left(\cdot\right)$ is the link function, $\mu_{ap} \equiv \E\left[y_{ap}\right]$ is equivalent to the expected value of the response and $f_A$, $f_P$ and $f_C$ are the smooth functions of age $a$, period $p$ and cohort $c$. The APC identification problem means we can add a constant and linear trend to each function without affecting the overall linear predictor. Consider the following functions \citep{Carstensen2007age}
\begin{align*}
	\tilde{f}_A\left(a\right) & = f_A\left(a\right) -  \constone + \constthree{M}{a} \\
	\tilde{f}_P\left(p\right) & = f_P\left(p\right) + \constone + \consttwo - \constthree{p} \\
	\tilde{f}_C\left(c\right) & = f_C\left(c\right) - \consttwo + \constthree{c}
\end{align*}
where $\constone$ and $\consttwo$ are due to the identification for including more than one smooth function and $\constthree$ is due to the structural link. The overall linear predictor is invariant to the inclusion of these constants since
\begin{align*}
	g & \left(\tilde{\mu}_{ap}\right) \\
	& =  \tilde{f}_A \left(a\right) + \tilde{f}_P\left(p\right) + \tilde{f}_C\left(c\right)  \\
	\begin{split}
		& =  \left[f_A\left(a\right) -  \constone + \constthree{M}{a}\right] + \\
		& \qquad \left[f_P\left(p\right) + \constone + \consttwo - \constthree{p}\right] + \\
		& \qquad \left[f_C\left(c\right) - \consttwo + \constthree{c}\right]
	\end{split} \\
	& =  \left[f_A\left(a\right) + \constthree{M}{a}\right] + \left[f_P\left(p\right) - \constthree{p}\right] + \left[f_C\left(c\right) + \constthree{c}\right] \\
	& =  f_A \left(a\right) + f_P\left(p\right) + f_C\left(c\right) \\
	& = g\left(\mu_{ap}\right)
\end{align*}
where $c = p - M \times a$ is used in the fourth equality.

Many re-parameterisation schemes are based off of an identifiable set of quantities. The first derivatives are not identifiable \citep{Mason1973, fienberg1979identification} but the second derivatives, or more generally curvatures, are identifiable. The age curvature is expressed
\begin{equation*}
	\tilde{f}_{A_C}\left(a\right) \equiv \tilde{f}''_A\left(a\right) = f''_A\left(a\right) \equiv f_{A_C}\left(a\right)
\end{equation*}
where the subscript ``$C$'' denotes the curvature. The terms for period and cohort curvature are analogous. 

\subsection{Univariate temporal model}

For the purpose of development, we shall first focus on, a single temporal function for age. A univariate temporal model for age can be expressed as
\begin{equation}
	\label{Eq: age-model: smooth only}
	g\left(\mu_a\right) = f_A\left(a\right)
\end{equation}
for $f_A$, a smooth function of covariate $a$ for age.

A popular set of functions used to approximate the smooth functions are splines: sums of polynomial functions called basis functions, which are based on a selection of points called knots. Within APC modelling, the \texttt{Epi} \citep{Carstensen2007age} package in \texttt{R} \citep{Rproject} fits several splines bases to continuous functions of APC models without penalisation. Carstensen also incorporated his methods into a package in \texttt{Stata} with extensions to include covariates. \citep{Rutherford2010}

To approximate $f_A$ in Eq.\eqref{Eq: age-model: smooth only}, the user specifies basis functions, and the model fitting process produces estimates for the weights of said basis functions. Given the basis $b_i\left(a\right)$, the $i^\text{th}$ basis function, $f_A$ is approximated with a spline as follows
\begin{equation*}
	\hat{f}_A \left(a\right) = \sum_{i=1}^{I}b_i\left(a\right)\hat{\beta}_i	
\end{equation*}
where $I$ is the number of basis functions and $\hat{\beta}_i$ is the estimate of the unknown weights.

Estimates of the true function can be found using a penalised iterative re-weighted least squares (PIRLS) algorithm to produce $\hat{\beta}_i$. \citep{Wood2017generalized} PIRLS is used to find an estimate $\hat{\bs{f}}_A$ that minimises the objective function
\begin{equation*}
	\label{Eq: PIRLS minimisation}
	D\left(f_A\left(a\right)\right) + \lambda_A \int f_A''\left(a\right)^2 da
\end{equation*}
where $D\left(f_A\left(a\right)\right)$ is the deviance (square of the difference between the saturated log-likelihood and model log-likelihood) of the model and $\lambda_A \int f_A''\left(a\right)^2 da$ is a penalty term on the second derivative ``wiggliness'' of $f_A$ with smoothing parameter $\lambda_A$. For more details, see Chapter 4 Wood (2017). \cite{Wood2017generalized} By representing the smooth function via a spline basis, the smooth itself can be written
\begin{equation*}
	\label{Eq: age-model: smooth only - Matrix}
	f_A\left(a\right) = \sum_{i=1}^{I}b_i\left(a\right)\beta_i = \bs{X}\bs{\beta}
\end{equation*}
for $\bs{X}$ an $n \times I$ matrix and $\bs{\beta}$ an $I \times 1$ vector of parameters. The penalty function can be expressed as
\begin{equation*}
	\label{Eq: age-model: penalty function as matrix}
	\int f_A''\left(a\right)^2 da = \bs{\beta}^T \int \bs{b}^T \left(a\right) \bs{b} \left(a\right) da \bs{\beta} = \bs{\beta}^T \bs{S}_A \bs{\beta}
\end{equation*}
where $\bs{S}_A = \int \bs{b}^T \left(a\right) \bs{b} \left(a\right) da$ is the penalty matrix.

Penalising estimates of the smooth function reduces the effect of over fitting (e.g., from choosing too many bases to represent the smooth function) as over-fit functions are often ``wigglier'' than those under-fit and hence penalised greater. The smoothing parameter controls the trade-off between smoothness of estimated smooth and closeness to the data. If $\lambda_A = 0$, there is no cost for fitting complicated functions while $\lambda_A \rightarrow \infty$ gives the maximum cost for fitting a complicated function, and $\hat{f}_A$ is a straight line.

\subsection{Orthogonalization}

Often an intercept is included alongside smoothers; this causes identifiability problems that can be resolved via re-parameterisation. A `sum-to-zero' constraint orthogonalizes the smooth to an intercept term such that $\bs{1}^T\bs{X}\bs{\beta}=\bs{0}$, avoiding any intercept related identification problems. The constraint is applied by constructing an $I \times \left(I-1\right)$ matrix $\bs{Z}$ through the QR-decomposition of $\left(\bs{1}^T\bs{X}\right)^T$. The smooth is re-parameterised by using $\bs{X}\bs{Z}$ and $\bs{Z}^T\bs{S}\bs{Z}$ as its model and penalty matrices; for more details, see Chapter 5 Wood (2017).\citep{Wood2017generalized}

The parameter space of $f_A$ can be split further into a linear slope and parameters corresponding to orthogonal curvatures. \citep{holford1983estimation} In the same vein as the intercept re-parameterisation, define a $2 \times I$ array consisting of a constant and vector of all ages for the intercept and linear terms, $\left[ \bs{1} : \bs{a} \right]$. Consequently, a $\left(I-1\right) \times \left(I-2\right)$ matrix $\bs{Z}$ is calculated by the QR-decomposition of
$\left(\left[ \bs{1} : \bs{a} \right]^T \bs{X}\right)^T$, and the smooth $f_A$ is re-parameterised using $\bs{A}_C = \bs{X}\bs{Z}$ and $\bs{S}_{A_C} = \bs{Z}^T\bs{S}\bs{Z}$ as its model and penalty matrices. The subscript $C$ denotes ``curvature''.

After the intercept and linear slope re-parameterisation, the form of the age-model is
\begin{equation*}	
	\label{Eq: age-model: intercept, linear and smooth}
	g\left(\mu_a\right) = \beta_0 + a\beta_{A_L} + f_{A_C}\left(a\right)
\end{equation*}
where $\beta_0$ and $\beta_{A_L}$ are the parameters for the intercept and slope and $f_{A_C}$ is the smooth of covariate $a$ orthogonal to the intercept and linear term. In matrix form,
\begin{equation*}
	g\left(\bs{\mu}\right) = \bs{X}\bs{\beta} = \left[\bs{1} : \bs{a} : \bs{A}_C\right]
	\left[
	\begin{array}{c}
		\beta_0 \\
		\beta_{A_L} \\
		\bs{\beta}^T_{A_C}
	\end{array}
	\right]
\end{equation*}
where $\beta_0$, $\beta_{A_L}$ and $\bs{\beta}_{A_C}$ are the parameters for the intercept, slope and curvature terms defined by the partitions $\bs{1}$, $\bs{a}$ and $\bs{A}_C$ of the model matrix. The smooth function $f_{A_C}$ has the associated penalty $\lambda_A \bs{\beta}^T_{A_C} \bs{S}_{A_C} \bs{\beta}_{A_C}$.

\begin{figure}[!h]
	\caption{Thin plate regression spline basis before any re-parameterisation, after an intercept re-parameterisation and after an intercept and slope re-parameterisation.}
	\label{Fig: smooth basis after constraints}
	\centering
	{\includegraphics[width=.75\textwidth]
	{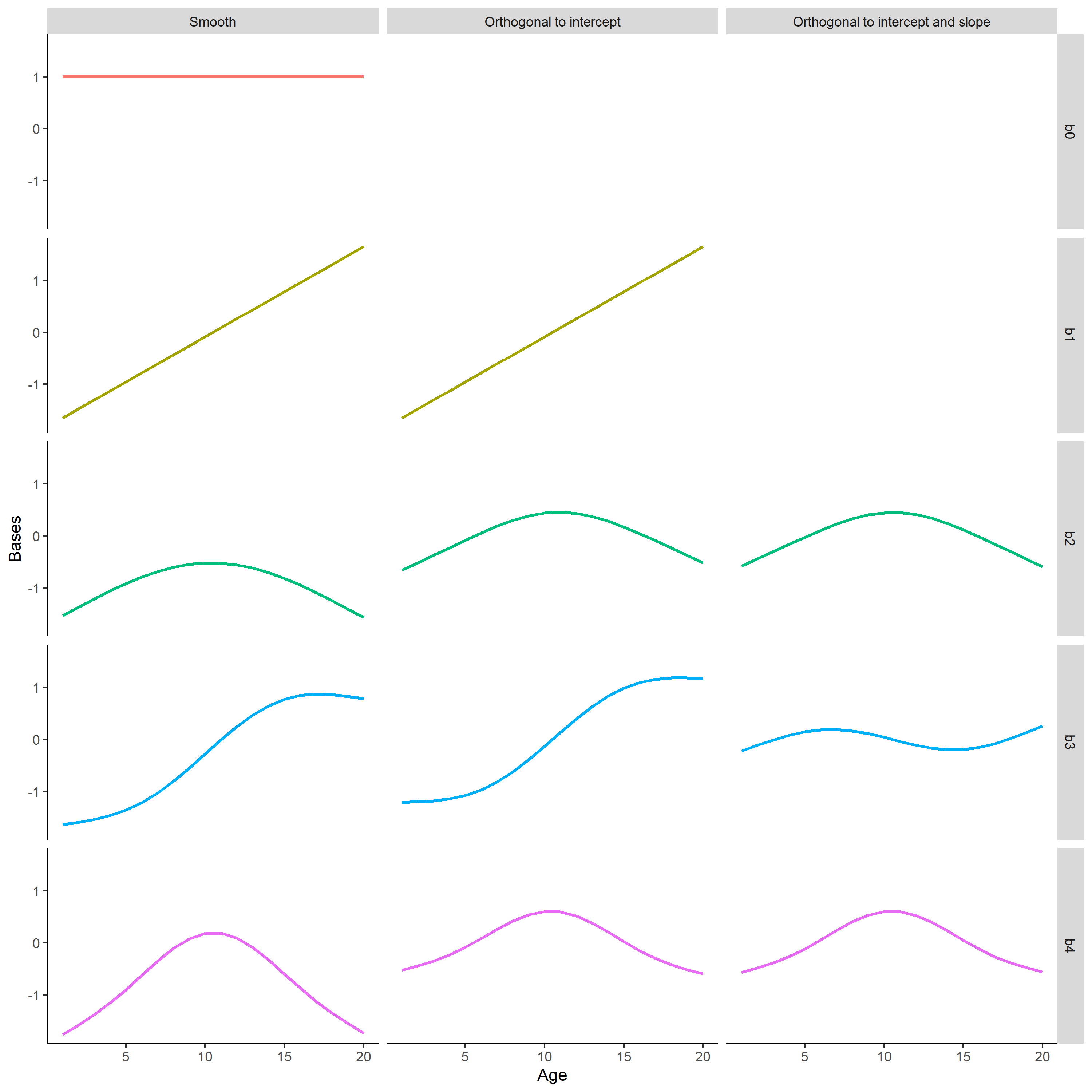}}
\end{figure}

Figure \ref{Fig: smooth basis after constraints} shows how a spline basis for $a=1,\dots,20$ with $i=1,\dots,5$ basis functions changes after each re-parameterisation. Each row relates to a basis function; $\texttt{b0}-\text{degree}(0)$, $\texttt{b1}-\text{degree}(1)$, $\texttt{b2}-\text{degree}(2)$, $\texttt{b3}-\text{degree}(3)$ and $\texttt{b4}-\text{degree}(4)$. More details on specification of the spline basis will follow at the end of this section. The first column is the basis without any re-parameterisation, the second is after being re-parameterised with respect to an intercept and the third is re-parameterised with respect to an intercept and a linear slope.
As can be seen in Figure \ref{Fig: smooth basis after constraints}, orthogonalization not only removes the bases corresponding to the desired constraint(s), it changes the overall shape of the remaining bases, see $\texttt{b3}$. Intercept orthogonalization orthogonalizes the smooth to a constant, $\text{degree}(0)$ - \texttt{b0} is removed. Intercept and slope orthogonalization orthogonalizes the smooth to a constant and a linear trend, $\text{degree}(0)$ and $\text{degree}(1)$ -  \texttt{b0} and \texttt{b1} are removed.

\subsection{Age-period-cohort modelling}
\label{Subsection: Reparameterised APC model}

After re-parameterising period and cohort in the same manner as age, an estimable APC model is written,
\begin{equation}
	\label{Eq: age-period-cohort model: full re-parameterisation}
	g\left(\mu_{ap}\right) = \beta_0 + s_1\beta_{1} + s_2\beta_{2} + f_{A_C}\left(a\right) + f_{P_C}\left(p\right) + f_{C_C}\left(c\right)
\end{equation}
where $s_1$ and $s_2$ are two of the three temporal slopes with parameters $\beta_{1}$ and $\beta_{2}$ and $f_{A_C}\left(a\right)$, $f_{P_C}\left(p\right)$ and $f_{C_C}\left(c\right)$ are the smooths for the age, period and cohort curvatures. If all three slopes are included in the above re-parameterisation, the model is over-parameterised. By dropping any one of the three slopes, the model is no longer over-parameterised, and the scheme is based off the identifiable curvatures that are invariant to the choice of slope dropped. \citep{holford1983estimation} That is, dropping any of the age, period or cohort slopes does not change the estimates of curvatures in any way.

To generalise what parameters are estimable, let $s_{a_L}$, $s_{p_L}$ and $s_{c_L}$ be the respective age, period, and cohort slopes. Any linear combination of $\beta_1 s_{a_L} + \beta_2 s_{b_L} + \left(\beta_2 - \beta_1\right) s_{c_L}$ is estimable for arbitrary $\beta_1$ and $\beta_2$. \citep{holford1983estimation} While individual slopes cannot be estimated, the recommendation from Holford is to drop one slope as the effect of the dropped slope is included in the remaining two, which is \textit{ad-hoc}.

Re-parameterisations using curvatures orthogonal to linear slopes, \citep{holford1983estimation} second differences \citep{kuang2008a} (second differences are the discrete way to define local curvature, like the continuous second derivatives) and period and cohort terms orthogonal to linear trends \citep{Carstensen2007age} provide systematic solutions to the APC problem. In each scheme, an arbitrary choice is made: which linear slope to drop, \citep{holford1983estimation} which three baseline rates to choose \citep{kuang2008a} and which term and reference term to use. \citep{Carstensen2007age} Consequently, we refer to the aforementioned schemes as `overall non-arbitrary' - they are based on a set of identifiable quantities but require an arbitrary choice during the re-parameterisation process.

Depending on which of the two slopes are kept in, the interpretation of the model changes. Commonly the age slope is often retained due to age's importance in most health concerns. If the cohort slope is dropped, the APC model is ``cross-sectional'', i.e.,
\begin{equation*}
	g\left(\mu_{ap}\right) = \beta_0 + a\beta_{1} + p\beta_{2} + f_{A_C}\left(a\right) + f_{P_C}\left(p\right) + f_{C_C}\left(c\right)
\end{equation*}
and if the period slope is dropped, the APC model is ``longitudinal'', i.e.,
\begin{equation*}
	g\left(\mu_{ap}\right) = \beta_0 + a\beta_{1} + c\beta_{2} + f_{A_C}\left(a\right) + f_{P_C}\left(p\right) + f_{C_C}\left(c\right).
\end{equation*}
In the remainder of the paper, we do not concern ourselves with the interpretation of the model based off of the slope dropped and choose to drop the cohort slope in all subsequent models for consistency. 

For models with multiple smooth functions, the penalty in the objective function is the addition of each individual smooths penalty function. For APC models re-parameterised as above, the penalty function is
\begin{equation*}
	\lambda_a\int_a f_{A_C}''\left(a\right)^2da +
	\lambda_p\int_p f_{P_C}''\left(p\right)^2dp +
	\lambda_c\int_c f_{C_C}''\left(c\right)^2dc,
\end{equation*}
or in matrix form,
\begin{equation*}
	\lambda_a\bs{\beta}_{A_C}\bs{S}_{A_C}\bs{\beta}_{A_C} +
	\lambda_p\bs{\beta}_{C_C}\bs{S}_{P_C}\bs{\beta}_{P_C} +
	\lambda_c\bs{\beta}_{P_C}\bs{S}_{C_C}\bs{\beta}_{C_C}.
\end{equation*}

\subsection{Implementation}

There are many types of spline basis functions one might use with common choices including thin plate regression splines (TPRS) and cubic regression splines (CRS). TPRS smooth with respect to any number of covariates and do not need the knots to be specified \textit{a priori}; however, TPRS are computationally costly and are not invariant to rescaling of the covariate. CRS are computationally cheap with directly interpretable parameters but can only model one covariate at a time and require the knots to be predefined. For more details on these bases and examples of others, see Chapter 5 Wood (2017). \citep{Wood2017generalized} The example in Figure \ref{Fig: smooth basis after constraints} uses TPRS basis for illustration purposes but we will use a CRS basis in the remainder of this paper.

The APC model in Eq.\eqref{Eq: age-period-cohort model: full re-parameterisation} has a parametric component, the two included slopes, and a non-parametric component, the smooth functions of curvatures. Therefore, it can fit into the wider framework of a generalised additive model (GAM). \citep{Hastie1990} We implement the GAMs in the \texttt{mgcv} package \citep{Wood2017generalized} in \texttt{R} \citep{Rproject} which offers a wide range of spline bases to represent smooth functions and their penalties. An example formula for a GAM in \texttt{mgcv} with one parameteric component (\texttt{x1}) and two non-parameteric components (\texttt{x2} and \texttt{x3}) represented by CRS is $\texttt{y } \sim \texttt{ x1 } + \texttt{ s(x2, bs=``cr'', k=x2k) } + \texttt{ s(x3, bs=``cr'', k=x3k) } $. Here \texttt{s()} is the call to a univariate smooth, \texttt{bs} is the argument to specify the basis to use (e.g. ``\texttt{cr}'' for CRS) and \texttt{k} is the basis dimensions (the knots in the case of a spline). In order to manually modify the model and penalty matrices, we utilise the $\texttt{fit = FALSE}$ argument in $\texttt{gam()}$. This arguments returns the full model, including model and penalties matrices, before the model fitting process. We take model and penalty matrices returned here, modify them as required, and then perform the model fitting process. Code to replicate the following simulation studies and application analysis can be found at \url{https://github.com/connorgascoigne/Unequal-Interval-APC-Models}.

\section{Simulation Study}
\label{Section: Simulation Study}

We now present a simulation study to demonstrate that the proposed model provides a suitable solution to the APC identification problem in the simplest, $M=1$, scenario.

\subsection{Data}

The simulation study is motivated by obesity rates and how they increase with age and in more recent years, \citep{Flegal2002, Mokdad2003, Ogden2006} whilst also having an hereditary effect. \citep{Cole2008, Gillman2004} Obesity data is a common application of APC models as a range of different responses can  be used. For example, a linear regression model can be used on weights. \citep{Luo2015} Alternatively, body mass index (BMI) has been modelled via a linear regression model (using log-BMI) and a logistic regression model (indicator of a given BMI). \citep{Fannon2021} In addition, a Poisson model for counts of rare events can be used.

The shapes for the age, period and cohort effects are adopted from a simulation study for a Gaussian data from Luo and Hodges. \citep{Luo2015} We extend this study to include responses from binomial and Poisson paradigms. Specific choices for the simulation set up and distribution parameters in the binomial and Poisson cases are motivated by the UKs yearly obesity survey from the National Health Service (NHS). \citep{NHS2020} The survey is of approximately 8000 adults grouped from 16-24 up to 75+. 

Data is simulated for individuals in single-year age-period format. The time index is continuous with the yearly midpoint being taken for the modelled value. We define single-year ages between $\left[0,60\right]$ and single-year period between $\left[0, 20\right]$. Cohort is found using $c=p-a$. For the normal and Binomial distributions, $N_{ap}$ reflects the number of individuals included in the survey which for each of the 60 ages is fixed to be 150. For the Poisson distribution, $N_{ap}$ is typically the population at risk, but for consistency this will be kept at 150 as well.

The true functions of age, period and cohort to generated the Gaussian data are
\begin{equation*}
	\label{Eq: True functions}
	\begin{split}
		h_A\left(a\right) & = 0.3a - 0.01a^2 \\
		h_P\left(p\right) & = -0.04p +0.02p^2 \\
		h_C\left(c\right) & = 0.35c -0.0015c^2.
	\end{split}
\end{equation*}
In order to use the same set of $h_\star$ functions (so the curve shapes are consistent across distributions), the simulations for the binomial and Poisson case are altered via an offset and scaling factor
\begin{equation*}
	\text{offset} + \text{scale} \times \left[h_A\left(a\right) + h_P\left(p\right) + h_C\left(c\right)\right]
\end{equation*}
to match the obesity survey data. The expected responses for the binomial (overweight, $\text{BMI} \geq 25$) and Poisson (obese, $\text{BMI} \geq 30$) data reflect an average of approximately 64\% and 28\% of the UKs adult population, respectively. Furthermore, both sets of responses have approximately 20\% difference between the age group with the smallest number of counts and largest. 

\sloppy The data from each distribution is generated from
\begin{align*}
	y_{nap}^s & \sim \text{Normal}\left(\mu_{ap}, 1\right) \\
	y_{ap}^s & \sim \text{Binomial}\left(N_{ap}, \pi_{ap}\right) \\
	y_{ap}^s & \sim \text{Poisson}\left(\lambda_{ap}\right)
\end{align*}
where $\mu_{ap} = 0 + \left[{h_A\left(a\right) + h_P\left(p\right) + h_C\left(c\right)}\right]/{1}$ for $n = \{1, \dots, N_{ap} = 150\}$ and $s = \{1, \dots, S = 100\}$ for each simulation. For the binomial and Poisson distributions, $s$ is the same as for the Gaussian case and $\pi_{ap} = \text{expit}\left(0.4 + \left[{h_A\left(a\right) + h_P\left(p\right) + h_C\left(c\right)}\right]/{50}\right)$ and $\lambda_{ap} = N_{ap}\exp\left(-1.5 + \left[{h_A\left(a\right) + h_P\left(p\right) + h_C\left(c\right)}\right]/{50}\right)$. The range of binomial and Poisson responses are approximately 45\% to 81\% and 9\% to 51\%, respectively, which match the target percentages with $\pm$20\%.

In this paper we will only report on the results for binomial data generated. The results for the other distributions are in the Supplementary Material which can be found at \url{https://github.com/connorgascoigne/Unequal-Interval-APC-Models/blob/main/final_supplementary.pdf}. Furthermore, the Supplementary Material contains an example of data generated without all three temporal trends present (cohort is missing). This example highlights that the issues are due to the structural link within the data rather than the re-parameterisation we propose.

\subsection{Models}

To each of the $S$ data sets, we fit the following models:
\begin{enumerate}
	\item Factor (FA) Model: A factor version of an APC model is written $g\left(\mu_{ap}\right) = \beta_0 + \alpha_a + \tau_p + \gamma_c$ where $\beta_0$ is the overall level, $\alpha_a$, $\tau_p$ and $\gamma_c$ are the $a$, $p$ and $c$ levels of the age, period and cohort factors, respectively. The interpretation of these factors if there was no structural link identification would be relative risks (for example, for age it is the difference between the overall level and the $a^\text{th}$ age group). Due to the structural link, the factors are unidentifiable and cannot be interpreted as such; consequently, this model was originally re-parameterised into a set of linear trends and their orthogonal curvatures. \citep{holford1983estimation}.  The factor version of the re-parameterised APC model is,
	\begin{equation*}
		\label{Eq: FA model}
		g\left(\mu_{ap}\right) = \beta_0 + s_1\beta_{1} + s_2\beta_{2} + \alpha_{C_a} + \tau_{C_p} + \gamma_{C_p}
	\end{equation*}
	where $s_{1}$ and $s_{2}$ are the two temporal slopes kept with parameters $\beta_1$ and $\beta_2$ and $\bs{\alpha}_{C}$, $\bs{\tau}_{C}$ and $\bs{\gamma}_{C}$ are the factor curvature terms. This original re-parameterisation is used as a benchmark for comparison in the simulation study but is still widely used with summaries available from a user-friendly web tool from the National Cancer Institute (https://analysistools.cancer.gov/apc/). \citep{rosenberg2014web}
	
	\item Smoothing spline models: Detailed in Section \ref{Section: Method}, a re-parameterisation in the style of the FA model but on a continuous version of the APC model using smoothing splines on the curvatures
	\begin{equation*}
		\label{Eq: PSS Model}
		g\left(\mu_{ap}\right) = \beta_0 + s_{1}\beta_{1} + s_{2}\beta_{1} + f_{A_C}\left(a\right) + f_{P_C}\left(p\right) + f_{C_C}\left(c\right)
	\end{equation*}
	where $s_{1}$ and $s_{2}$ are the two temporal slopes kept with parameters $\beta_1$ and $\beta_2$ and $f_{A_C}\left(a\right)$, $f_{P_C}\left(p\right)$ and $f_{C_C}\left(c\right)$ are the smooth functions of curvature. The smooth functions are represented by cubic regression splines with the number of knots approximately 25\% of the number of unique data points for each temporal effect. 
	\begin{enumerate}
		\item Regression Smoothing spline (RSS): This is the smoothing spline model fit \textbf{without} penalisation; common to fit spline APC in this manner. In \texttt{mgcv}, smoothing penalties are applied by default but are removed using the option \texttt{fx=TRUE}. 
		\item Penalised smoothing spline (PSS): This is the smoothing spline model fit \textbf{with} penalisation. The importance of penalisation will become clear in Section \ref{Section: Unequal Intervals}.
	\end{enumerate}
\end{enumerate}

For all models, the \textit{ad-hoc} choice of what linear slope to drop will be cohort (cross-sectional). Therefore, the models will contain age and period slopes and curvatures for all three temporal effects.

\subsection{Results}

Identification issues due to the structural link are resolved by the \textit{ad-hoc} forcing of one of the slopes to be zero. Due to this, comparisons between $h_\star$ and $\hat{h}_\star$ are inappropriate as the true effects do not have a zero linear trend. In order to compare the two sets of quantities, both the true values and estimates need to be based off functions of means that are identifiable, rather than unidentifiable.

Thus we define modified true and estimated effects which take into consideration the intercept and structural link identifiability. In practise, first define the linear predictor for all APC combinations (including ones not present due to the structural link identification), then the modified true effects are calculated by subtracting the overall mean of the linear predictor from the marginal temporal effect of the linear predictor. For example, the true age effect for all three distributions is calculated,
\begin{equation*}
	\label{Eq: True Age Effects}
	h^{+}_A\left(a\right) = 
	\begin{cases}
		\frac{1}{PC} \sum_{p=1}^P \sum_{c=1}^C \mu_{apc} - \frac{1}{APC}  \sum_{a=1}^A\sum_{p=1}^P \sum_{c=1}^C \mu_{apc} & \text{Gaussian} \\
		\frac{1}{PC} \sum_{p=1}^P \sum_{c=1}^C \text{logit}\left(\pi_{apc}\right) - \frac{1}{APC}  \sum_{a=1}^A\sum_{p=1}^P \sum_{c=1}^C \text{logit}\left(\pi_{apc}\right) & \text{Binomial} \\
		\frac{1}{PC} \sum_{p=1}^P \sum_{c=1}^C \text{log}\left(\lambda_{apc}\right) - \frac{1}{APC}  \sum_{a=1}^A\sum_{p=1}^P \sum_{c=1}^C \text{log}\left(\lambda_{apc}\right) & \text{Poisson}
	\end{cases}.
\end{equation*}
The intercept identifiability is addressed in the true effects by subtracting the overall mean from the marginal of the linear predictor. As the structural link identifiability cannot be removed as with the intercept identifiability, it is consolidated into an `average effect' of the remaining two terms. To see this explicitly and without the loss of generality, consider the Gaussian case where $\texttt{offset} = 0$ and $\texttt{scale} = 1$,
\begin{equation*}
	\label{Eq: Age True Effects}
	\begin{split}
		h^{+}_A\left(a\right) & = \underbrace{ \frac{1}{PC} \sum_{p=1}^P \sum_{c=1}^C \mu_{apc} }_{\text{marginal age effect}} - \underbrace{ \frac{1}{APC}  \sum_{a=1}^A\sum_{p=1}^P \sum_{c=1}^C \mu_{apc}}_{\text{overall mean} } \\
		& = \frac{1}{PC} \sum_{p=1}^P \sum_{c=1}^C \left[h_A\left(a\right) + h_P\left(p\right) + h_C\left(c\right)\right] - \frac{1}{APC}  \sum_{a=1}^A\sum_{p=1}^P \sum_{c=1}^C \left[h_A\left(a\right) + h_P\left(p\right) + h_C\left(c\right)\right] \\
		& \propto h_A\left(a\right) + \underbrace{ \frac{1}{PC} \sum_{p=1}^P \sum_{c=1}^C \left[h_P\left(p\right) + h_C\left(c\right)\right] }_{\text{Average period/cohort effect}}
	\end{split}
\end{equation*}
where the linear trends in period and cohort are consolidated together into an average effect. For all distributions, the true age curvatures are found by de-trending the true effects
\begin{equation*}
	\label{Eq: Age True Curvatures}
	h^{+}_{A_C}\left(\bs{a}\right) = \left(\bs{I}_A - \bs{H}_A\right)h^{+}_{A}\left(\bs{a}\right) 
\end{equation*}
where $\bs{I}_A$ is an $A \times A$ identity matrix and $\bs{H}_A = \left[\bs{1}:\bs{a}\right] \left(\left[\bs{1}:\bs{a}\right]^T\left[\bs{1}:\bs{a}\right]\right)^{-1}\left[\bs{1}:\bs{a}\right]^T$ is the hat matrix for an ordinary least squares fit of the age effect. To define the estimated effects, use the estimated linear predictor for all APC combinations instead of the true linear predictor, $\hat{\mu}_{apc}$, $\text{logit}\left(\hat{\pi}_{apc}\right)$ and $\text{log}\left(\hat{\lambda}_{apc}\right)$ for Gaussian, binomial and Poisson, respectively, to define the marginal age effect and overall mean. The estimated curvatures are found using the estimated effects in the same manor as the true curvatures were. In all three distributions, the period and cohort effects and curvatures are analogous.

\begin{figure}[!h]
	\caption{Simulation study results for equal interval, $M=1$, binomial data generated when all three temporal effects are present. The FA, RSS and PSS models are the factor, regression smoothing spline and penalised smoothing spline models, respectfully. The first and second row are of the temporal effect and curvature plots for all models alongside the true values. The bottom two rows are the bias and MSE box plots for each model.} 
	\label{Fig: binom1APCResPlot}
	\centering
	{\includegraphics[width=.75\textwidth]
	{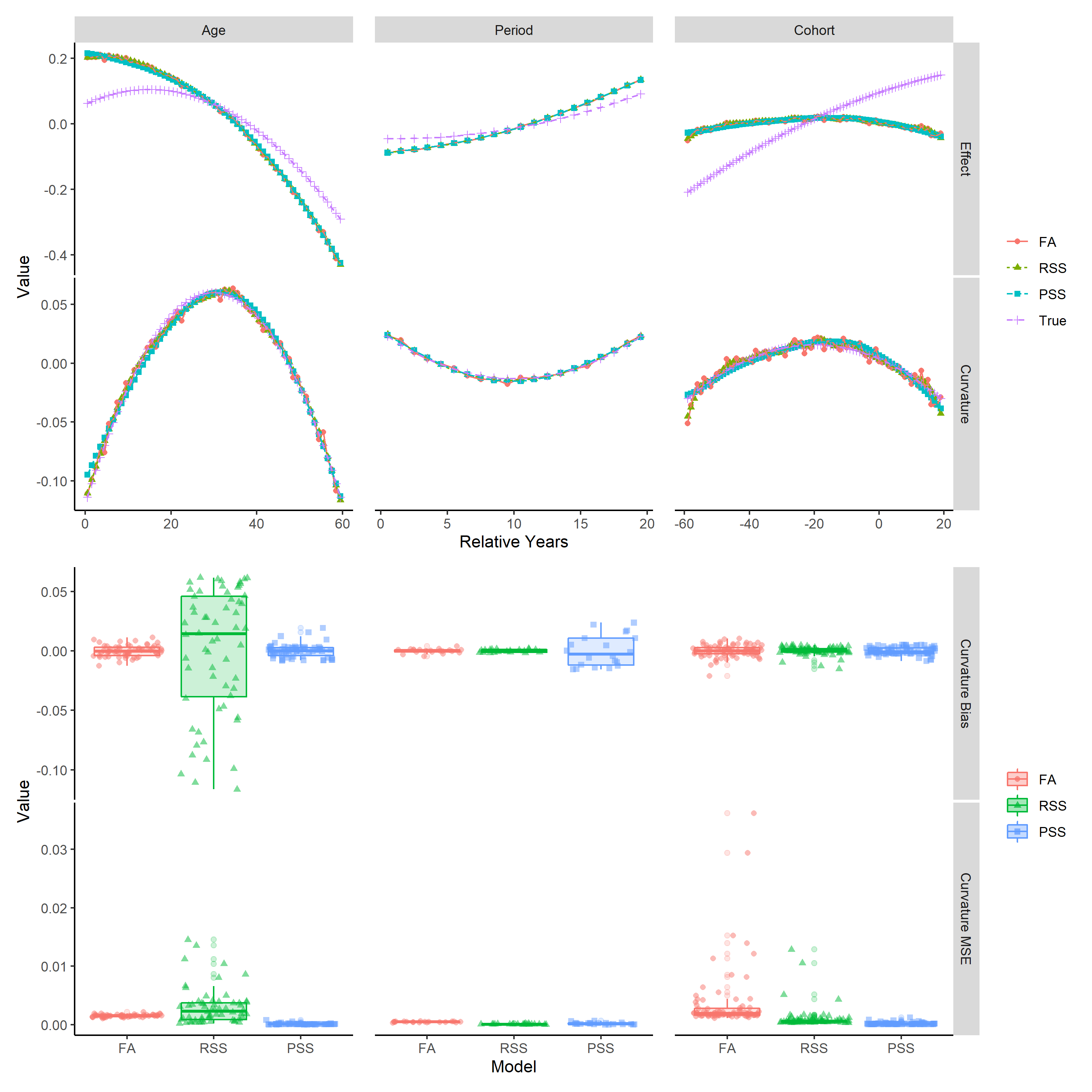}}
\end{figure}

The results of the binomial simulation study are summarised in Figure \ref{Fig: binom1APCResPlot}. Each column refers to one of the temporal effects; age, period and cohort from left to right. The first two rows show the estimated effect and curvature for each of age, period and cohort alongside their respective true effect and curvature. The latter two rows show the bias and mean square error (MSE) of the identifiable curvature terms. The bias and MSE for the age effect with $a=a$ are $\frac{1}{S}\left[\sum_{s=1}^{S} \left( \hat{h}^{+}_{A_s}\left(a\right) - h^{+}_{A}\left(a\right) \right) \right] $ and $\frac{1}{S}\left[\sum_{s=1}^{S}\left(\hat{h}^{+}_{A_s}\left(a\right) - h^{+}_{A}\left(a\right)\right)^2\right]$, respectively and is analogous for period and cohort and the curvatures.

The structural link means the individual linear trends of age, period and cohort cannot be found. In the re-parameterisation we forced the cohort trend to be zero, meaning the true linear trend in cohort will be present in the age and period linear terms. Similar can be said if the age or period linear trends are forced to be zero. The identification problem along with the \textit{ad-hoc} choice to resolve it is seen in the first row of Figure \ref{Fig: binom1APCResPlot}: the cohort model effects are relatively flat and do not match the true effects.

The identifiability of the curvatures is shown in the second row. The estimated curvatures for all models are like one another and like the true curvatures. The FA curvature estimates are not as smooth as those from the RSS and PSS models. This is due to the fact the RSS and PSS models smooth between terms with the PSS also penalising the ``wiggliness'' of the estimated functions. In addition, the added variability seen in the oldest and youngest cohorts (as these are the observations seen the least) is smoothed over in the RSS and PSS models and not in the FA model.

The bias and MSE are only displayed for the estimated curvatures as these are the identifiable component. Each model produces a set of curvatures that accurately estimate the true curvatures. The age bias for the RSS model is slightly larger than the other two but is still small ($\pm 0.05$). The MSE further highlights the adequateness of each model. The behaviour of the PSS model for data generated in equal intervals is consistent, if not outperforming, what is expect from the well-known and well used FA and RSS models.

\section{Unequally Aggregated Intervals For Age, Period and Cohort}
\label{Section: Unequal Intervals}

Temporal data aggregated into intervals that match (five-year age, five-year period) are referred to as in `equal intervals'. If they do not match (five-year age, single-year period), the data is referred to as in `unequal intervals’. Providers of health and demographic data frequently release data that has been aggregated over multiple years. Even if collected in single years, it is common to be released aggregated over multiple years, to preserve anonymity.

Unequally aggregated data can be considered in the simpler equal interval framework by collapsing over the lowest common multiple (LCM) of the intervals, LCM$\left(\text{age-years}, \text{period-years}\right)$. Consider the following two cases LCM$\left(2,1\right)=2$ and LCM$\left(5,3\right)=15$. In the former, period is collapsed over two-groups leading to some information loss but potentially removes noise that obscures the true trend. In the latter, age is collapsed over three- and period over five-groups resulting in a larger amount of information lost. The more groups collapsed over, the fewer observations there are, inducing greater uncertainty in the parameter estimates.

\subsection{Added identification problems}

Previously we have focused on the case where age and period are in equal intervals, $M=1$. Table \ref{Tab: M=5 ap data table} shows how the cohort index varies when age is aggregated into an interval five-times larger than period, $M=5$. Cohorts appear every fifth period, highlighted in blue. Due to this, you can add a constant to the effect of every $M^\text{th}$ period (i.e., $1,6,11,\dots$) and subtract the same constant to the effect of every $M^\text{th}$ cohort (i.e., $1,6,11,\dots$) without changing the linear predictor.
\begin{table}[!h]
	\caption{Cohort indexing for age-period data table where age is grouped $M=5$ times larger than period. The cohort index is defined using $c=M\times\left(A-a\right)+p$ where $A=8$ to fix the first cohort to be 1.}
	\label{Tab: M=5 ap data table}
	\centering
	\begin{tabular}{ccccccccccc}
		\hlinewd{2pt} 
		8 & \cellcolor{blue!25}{1} & 2 & 3 & 4 & 5 & \cellcolor{blue!25}{6} & 7 & 8 & 9 & 10 \\
		7 & \cellcolor{blue!25}{6} & 7 & 8 & 9 & 10 & \cellcolor{blue!25}{11} & 12 & 13 & 14 & 15 \\ 
		6 & \cellcolor{blue!25}{11} & 12 & 13 & 14 & 15 & \cellcolor{blue!25}{16} & 17 & 18 & 19 & 20 \\ 
		5 & \cellcolor{blue!25}{16} & 17 & 18 & 19 & 20 & \cellcolor{blue!25}{21} & 22 & 23 & 24 & 25 \\
		4 & \cellcolor{blue!25}{21} & 22 & 23 & 24 & 25 & \cellcolor{blue!25}{26} & 27 & 28 & 29 & 30 \\
		3 & \cellcolor{blue!25}{26} & 27 & 28 & 29 & 30 & \cellcolor{blue!25}{31} & 32 & 33 & 34 & 35 \\
		2 & \cellcolor{blue!25}{31} & 32 & 33 & 34 & 35 & \cellcolor{blue!25}{36} & 37 & 38 & 39 & 40 \\   
		1 & \cellcolor{blue!25}{36} & 37 & 38 & 39 & 40 & 41 & 42 & 43 & 44 & 45 \\ 
		\hlinewd{1.5pt}
		\textbf{Age} & 1 & 2 & 3 & 4 & 5 & 6 & 7 & 8 & 9 & 10 \\
		\cline{2-11}
		& \multicolumn{10}{c}{\textbf{Period}} \\
		\hlinewd{2pt}
	\end{tabular}
\end{table}

As well as the identification problems for equal intervals, the additional identification issues that arise when modelling unequally aggregated APC data, $M \neq 1$, are encapsulated by the inclusion of two $M$ periodic functions $v_M\left(p\right) $ and $v_M\left(c\right)$ in a set of transformed functions. As $v_M$ is periodic, $v_M\left(x + M\right) = v_M\left(x\right)$ and the subscript is used to denote the periodicity. The transformed functions that include all the identification issues are
\begin{equation}
	\label{Eq: Unequal Transformed APC functions}
	\begin{split}
		\tilde{f}_A\left(a\right) & = f_A\left(a\right) -  \constone + \constthree{M}{a}, \\
		\tilde{f}_P\left(p\right) & = f_P\left(p\right) + v_M\left(p\right) + \constone + \consttwo - \constthree{p}, \\
		\tilde{f}_C\left(c\right) & = f_C\left(c\right) - v_M\left(c\right) - \consttwo + \constthree{c}.
	\end{split}
\end{equation}
As previously discussed, the overall linear predictor is invariant to the inclusion of a constant (intercept) and linear term (structural link). Without loss of generality let $\constone = \consttwo = \constthree = 0$, the linear predictor for the transformed functions is
\begin{align*}
	g & \left(\tilde{\mu}_{ap}\right) \\
	& =  \tilde{f}_A \left(a\right) + \tilde{f}_P\left(p\right) + \tilde{f}_C\left(c\right)  \\
	& =  f_A\left(a\right) + \left[f_P\left(p\right) + v_M\left(p\right)\right] + \left[f_C\left(c\right) - v_M\left(p-Ma\right)\right] \\
	& =  f_A\left(a\right) + \left[f_P\left(p\right) + v_M\left(p\right)\right] + \left[f_C\left(c\right) - v_M\left(p\right)\right] \\
	& =  f_A \left(a\right) + f_P\left(p\right) + f_C\left(c\right) \\
	& = g\left(\mu_{ap}\right)
\end{align*}
where $c=p-Ma$ and $v\left(m + M \right) = v\left(m\right)$ are used in the third and fourth lines, respectively. The linear predictor is invariant to the periodic function. 

Defining the second derivatives of the two periodic functions as 
\begin{equation*}
	\label{Eq: v_M curvatures}
	\begin{split}
		v''_M\left(p\right) & \equiv v_{M_C}\left(p\right) \\ 
		v''_M\left(c\right) & \equiv v_{M_C}\left(c\right)
	\end{split}
\end{equation*}
the previously identifiable second derivatives
\begin{equation}
	\label{Eq: Non-equal Curvature Terms}
	\begin{split}
		\tilde{f}''_{A}\left(a\right) \equiv \tilde{f}_{A_C}\left(a\right) & = f_{A_C}\left(a\right) \\ 
		\tilde{f}_{P_C}\left(p\right) & = f_{P_C}\left(p\right) + v_{M_C}\left(p\right) \\ 
		\tilde{f}_{C_C}\left(c\right) & = f_{C_C}\left(c\right) - v_{M_C}\left(c\right)
	\end{split}
\end{equation}
are no longer identifiable for period cohort due to the presence of $v_{M_C}$. The period and cohort curvature identifiability issues mean these terms are no-longer estimable, as in the equal interval case; in the estimate, we cannot disentangle what is the ``true'' curvature and was is the periodic function. We will call this the ``curvature identifiability problem''.

\subsection{Resolving the curvature identifiability problem}

Carstensen uses continuous functions to alleviate issues relating to the aggregation of years; \citep{Carstensen2007age} however, he does not explore the curvature identifiability problem from the unequal intervals and whether continuous functions alone resolve this. Holford recognised the curvature identifiability problem (calling it the micro-trend identifiability problem) and described how it can produce an $M$-year cyclic pattern in the estimated temporal effects as well as proposing the use of smooth functions to model them. \citep{holford2006approaches} He argued that smooth functions resolve the curvature identifiability problem by providing sufficient structure to smooth over the cyclic nature. This latter proposal has not been fully explored and only demonstrated as a solution for a real-world data example.

When fitting APC models to unequal data, the curvature identifiability problem means the period and cohort curvature functions are non-longer estimable Eq.\eqref{Eq: Non-equal Curvature Terms}. Because of this, an infinite number of period and cohort estimates can be used to produce the same re-parameterised linear predictor. Of those, if the estimate for the transformed period and cohort curvatures is estimating $v_M \neq 0$, the period and cohort estimates will display the, arbitrarily large, cyclic pattern over $M$-years as described by Holford. \citep{holford2006approaches} If $v_M \neq 0$ is estimated, the cyclic pattern in the period and cohort estimates is ``wigglier'' than when $v_M = 0$ is estimated. The PSS metod we proposed in Section \ref{Section: Method} has a penalty on the integrated square of the second derivative (``wiggliness'') of the estimates. Therefore, an estimate estimating $v_M \neq 0$ will have a larger integrated square of the second derivative and hence a larger penalty than an estimate estimating $v_M = 0$. Consequently, the PSS method we propose will actively penalise the curvature identification issues; whereas both Holford and Carstensen only smooth over them.

The re-parameterised APC model penalty function from Section \ref{Subsection: Reparameterised APC model} with the transformed functions is
\begin{equation*}
	\lambda_A \int \tilde{f}''_{A_C}\left(a\right)^2da + \lambda_P \int \tilde{f}''_{P_C}\left(p\right)^2dp + \lambda_C \int \tilde{f}''_{C_C}\left(c\right)^2dc 
\end{equation*}
where $\lambda_\star$ are the temporal smoothing parameters controlling the trade off between the estimates fit and smoothness. $\lambda_\star \rightarrow \infty$ and $\lambda_\star = 0$ lead to straight line and un-penalised estimates, respectfully. When the the integrated square of the second derivative is large, the smoothing parameter increases in order to applying additional cost to fitting the complicated function. Therefore, if the transformed period and cohort curvature estimates estimate $v_M \neq 0$, the larger integrated square of the second derivative, in comparison to when $v_M = 0$ is estimated, yields a greater cost to fit. 

Due to the curvature identifiability problem, we propose for fixed $\lambda_P$ and $\lambda_C$ the estimates of the transformed period and cohort functions with $v_M = 0$ have the smallest integrated squared second derivatives. This results in estimates for the transformed functions with $v_M = 0$ having the smallest penalty and 
\begin{equation*}
	\begin{split}
		\lambda_P \int \tilde{f}''_{P_C}\left(p\right)^2dp + \lambda_C \int \tilde{f}''_{C_C}\left(c\right)^2dc & = \lambda_P \int \left[f''_{P_C}\left(p\right)+v''_{M_C}\left(p\right)\right]^2dp + \lambda_C \int \left[f''_{C_C}\left(c\right)-v''_{M_C}\left(c\right)\right]^2dc \\ 
		& \geq \lambda_P \int f''_{P_C}\left(p\right)^2dp + \lambda_C \int f''_{C_C}\left(c\right)^2dc.
	\end{split}	
\end{equation*}
When estimating APC trends for unequal data, the age curvature is still identifiable Eq.\eqref{Eq: Non-equal Curvature Terms}. Consequently, and for the purpose of explanation, we omit the age curvature penalty term from the above expression and in the below theoretical section. However, we would like to stress that the age curvature penalty is still present in the estimation process.

To demonstrate the penalty term for period and cohort curvature estimates is greater when $v_M \neq 0$ is estimated than for when $v_M = 0$ is estimated, consider the special case using a natural cubic spline with knots spaced $M$ apart. Natural cubic splines enforce linearity beyond the boundary knots which reduces instability in these regions. This is a useful quality for the likes of cohort where the tails are based on few observations. \citep{heuer1997modeling} First, we consider period.

Both $\tilde{f}_{P_C} $ and $f_{P_C}$ are defined by the same set of knots $\check{p}_1 < \dots <\check{p}_J$ which are spaced $M$ apart from one another. The natural cubic splines are continuous to the second derivatives ($f'''_{P_C}$ is constant in and $\check{p}_j^{+}$ is a point in $\left[\check{p}_j, \check{p}_{j+1}\right]$) and are linear beyond the boundary knots ($f''_{P_C}\left(\check{p}_1\right)=f''_{P_C}\left(\check{p}_J\right)=0$). Therefore, the penalty term of $\tilde{f}_{P_C}$ can be written as
\begin{equation*}
	\int_{\check{p}_1}^{\check{p}_J} \tilde{f}''_{P_C}\left(p\right)^2 dp = \int_{\check{p}_1}^{\check{p}_J} f''_{P_C}\left(p\right)^2 dp + 2 \int_{\check{p}_1}^{\check{p}_J} f''_{P_C}\left(p\right)v''_{M_C}\left(p\right)dp + \int_{\check{p}_1}^{\check{p}_J} v''_{M_C}\left(p\right)^2 dp
\end{equation*}
when integrated over the range of all knots.

Applying integration by parts to the cross term
\begin{align*}
	\int_{\check{p}_1}^{\check{p}_J} f''_{P_C}\left(p\right)v''_{M_C}\left(p\right)dp & = \left. f''_{P_C}\left(p\right)v'_{M_C}\left(p\right) \right\rvert_{\check{p}_1}^{\check{p}_J} -  \int_{\check{p}_1}^{\check{p}_J} f'''_{P_C}\left(p\right)v'_{M_C}\left(p\right)dp \\
	& = -  \int_{\check{p}_1}^{\check{p}_J} f'''_{P_C}\left(p\right)v'_{M_C}\left(p\right)dp \\
	& \qquad \text{ as } f''_{P_C}\left(\check{p}_1\right)=f''_{P_C}\left(\check{p}_J\right)=0 \\
	& =  - \sum_{j=1}^{J-1} f'''_{P_C}\left(\check{p}_j^{+}\right) \int_{\check{p}_j}^{\check{p}_{j+1}} v'_{M_C}\left(p\right)dp \\
	& \qquad \text{ as } f'''_{P_C} \text{ is constant and } \check{p}_j^{+} \text{ is a point in } \left[\check{p}_j, \check{p}_{j+1}\right] \\ 
	& = - \sum_{j=1}^{J-1} \text{constant} \left[v_{M_C}\left(\check{p}_{j+1}\right)- v_{M_C}\left(\check{p}_{j}\right)\right] \\
	& = 0
\end{align*} 
since $\check{p}_j$ and $\check{p}_{j+1}$ are $M$ apart ($v_{M_C}\left(\check{p}_{j+1}\right)= v_{M_C}\left(\check{p}_{j}\right)$). We have shown that
\begin{align*}
	\int_{\check{p}_1}^{\check{p}_J} \tilde{f}''_{P_C}\left(p\right)^2 dp & = \int_{\check{p}_1}^{\check{p}_J} f''_{P_C}\left(p\right)^2 dp + \int_{\check{p}_1}^{\check{p}_J} v''_{M_C}\left(p\right)^2 dp \\
	& \geq \int_{\check{p}_1}^{\check{p}_J} f''_{P_C}\left(p\right)^2 dp
\end{align*}
as $\int_{\check{p}_1}^{\check{p}_J} v''_{M_C}\left(p\right)^2 dp \geq 0$.  

By applying the same reasoning as we did with period, we can similarly show that
\begin{align*}
	\int_{\check{c}_1}^{\check{c}_K} \tilde{f}''_{C_C}\left(c\right)^2 dc & = \int_{\check{c}_1}^{\check{c}_K} f''_{C_C}\left(c\right)^2 dc + \int_{\check{c}_1}^{\check{c}_K} v''_{M_C}\left(c\right)^2 dc \\
	& \geq \int_{\check{c}_1}^{\check{c}_K} f''_{C_C}\left(c\right)^2 dc.
\end{align*}
As a results, we have shown that when the curvature identifiability problem is present, the case when the estimate of the transformed period and cohort curvature functions is estimating $v_M \neq 0$, the integrated square of the second derivative of the transformed period and cohort curvature estimates is greater than when the estimates are estimating $v_M = 0$.

\subsection{Simulation study}

We now demonstrate this result emperically by repeating the simulaton study from Section \ref{Section: Simulation Study} with data in unequal intervals. We first generate the data in single-years and then aggregate, replicating the real-world practise of data being collected in single-year age and period format with aggregation occurring before the data is released. As is common in many epidemiology settings, we aggregate single-year age over five years (i.e., $M=5$).

The underlying single-year data are generated as described in Section \ref{Section: Simulation Study}. Once generated, the data is aggregated according to $\bs{p}$ and $\bs{a}'$, where $\bs{a}'$ is the $M$-year age vector of length $A'=A/M$. Once the aggregation has occurred, $\bs{p}$ and $\bs{a}'$ are used to define $\bs{c}'$, the cohort vector of length $C'= M\times\left(A'-1\right)+P$ using $c'=p-a'$.

In order to get a true age effect that is comparable to the estimated age effect, the true ages need to be averaged over every $M$ distinct ages. Let $a_i$ and $a'_i$ be the $i^\text{th}$ value in the vectors $\bs{a}$ and $\bs{a}'$, the true value of the age effect that is comparable to the aggregated estimated values is
\begin{equation*}
	h^{+}\left(a'_i\right) = \frac{1}{M} \sum_{m=-\left(M-1\right)}^{0} h^{+}\left(a_{\left[\left(i \times M\right) + m\right]}\right)
\end{equation*}
for $i = \{1, \dots, A'=\frac{A}{M}\}$. For example, we average the true age effect evaluated at 0.5, 1.5, 2.5, 3.5 and 4.5 in order to be comparable to the estimated effect at $a'=2.5$.

Similarly, for cohort, average over every $M$ cohorts (as age is aggregated in $M$ years) and move along in single-year steps (as period is still single years). Therefore, let $c_k$ and $c'_k$ be the $k^\text{th}$ value in the vectors $\bs{c}$ and $\bs{c'}$. The true value of the cohort effect that is comparable to the aggregated fitted values is
\begin{equation*}
	h^{+}\left(c'_k\right) =\frac{1}{M} \sum_{m=0}^{M-1} h^{+}\left(c_{k+m}\right)
\end{equation*}
where $k=\{1, \dots, C'=M\times\left(A'-1\right)+P\}$. For example, average the cohort true effects at $c = $-59, -58, -57, -56 and -55 to be comparable to the estimated effect at $c'=-57$.

Once the age and cohort true values are aggregated, the bias and MSE will reflect the variability observed in the $M=1$ simulations as well as the aggregation bias. As period is not changed, the expressions from Section \ref{Section: Simulation Study} are used. The models fit are the factor (FA), regression smoothing spline (RSS) and penalised smoothing spline (PSS) defined in Section \ref{Section: Simulation Study}. The estimated effects $\hat{h}^+_\star$ and curvatures $\hat{h}^+_{\star_C}$ are calculated like Section \ref{Section: Simulation Study} but with the vectors $a'$ and $c'$ for age and cohort; period is unchanged.

\begin{figure}[!h]
	\caption{Simulation study results for unequal interval, $M=5$, binomial data generated when all three temporal effects are present. The FA, RSS and PSS models are the factor, regression smoothing spline and penalised smoothing spline models, respectfully. The first and second row are of the temporal effect and curvature plots for all models alongside the true values. The bottom two rows are the bias and MSE box plots for each model.}
	\label{Fig: binom5APCResPlot}
	\centering
	{\includegraphics[width=.75\textwidth]
	{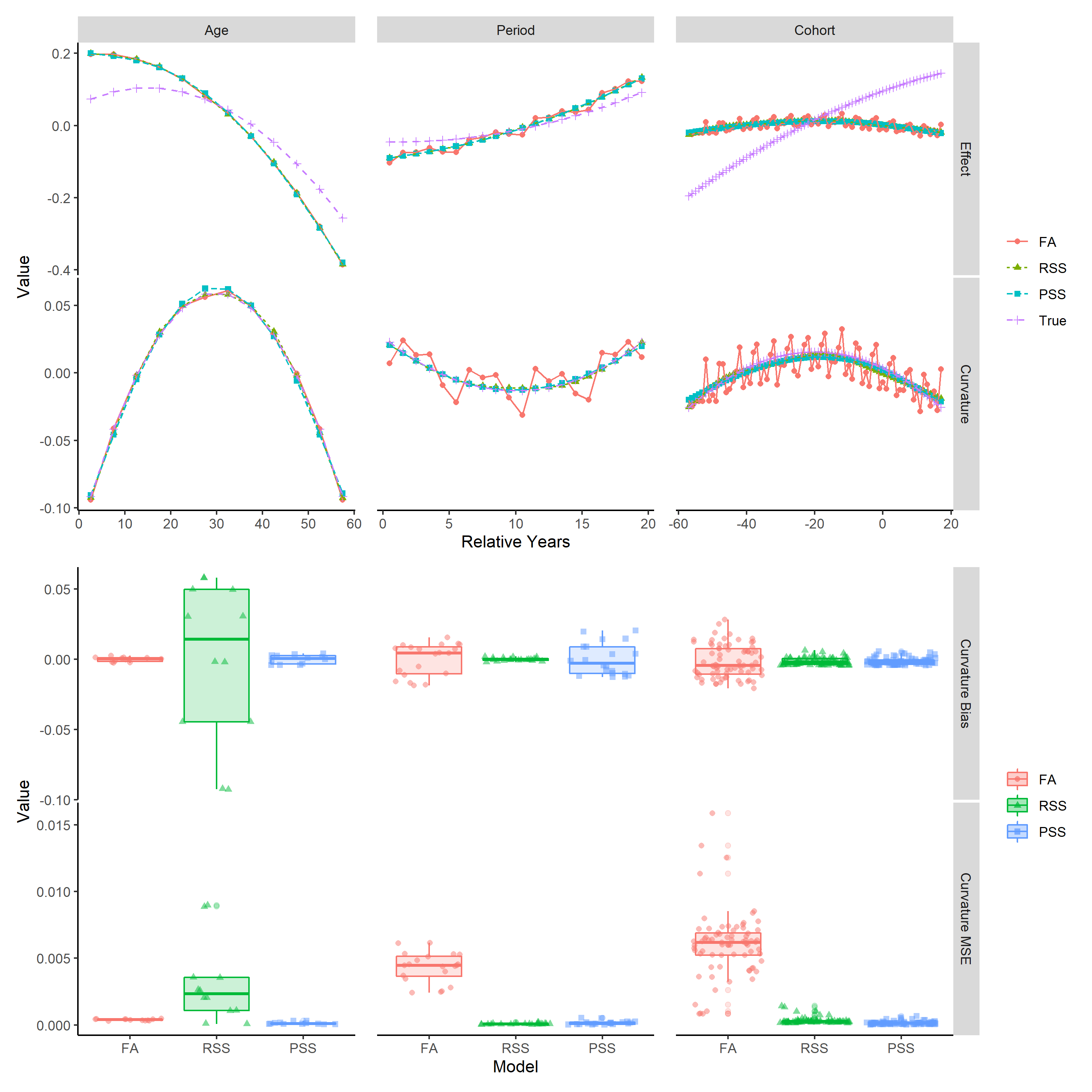}}
\end{figure}

Figure \ref{Fig: binom5APCResPlot} shows the results of the simulation study. Each column is one of the three temporal effects, the first two rows are the function plots of the estimated full effects and curvatures alongside the true functions of both and the bottom two rows are the bias and MSE for the curvatures.

The FA model displays the cyclic saw-tooth pattern which repeats every $M$-years (five-years) in both the full effects and curvatures for period and cohort, not age. The cyclic pattern in the period and cohort curvatures is due to the curvature identifiability problem. The period and cohort bias plots for FA model seem reasonable but this is due to the fact the cyclic pattern negating the overall bias. More telling is the difference between the FA models period and cohort MSE box plots and those from the RSS and PSS models; the FA box plot displays a general increase in the MSE values which themselves are more over-dispersed.

It is hard to differentiate between the results for the RSS and PSS models, with both seeming to provide adequate solutions to resolving the curvature identifiability problem. From the theoretical results, period and cohort curvature functions are not estimable, and when $v_M = 0$, the estimates of the trasformed period and cohort curvature functions are the smoothest. Therefore, the closeness between the RSS and PSS model estimates could be due to the fact that the CRS spline is estimating $v_M = 0$, which means the estimates are the smoothest and there is no additional penalty being incurred from the added identification.

To understand if the closeness between the RSS and PSS results is in fact due to the chosen basis function estimating $v_M = 0$ rather than $v_M \neq 0$, we perform an additional simulation study with two richer bases. A richer basis will offer another, of the infinite, choice for $v_M$ and given the basis is richer, it will be more likely that the estimate will be estimating $v_M \neq 0$. With a richer bases where $v_M \neq 0$, the curvature identifiability problem will be more apparent as will the difference between the RSS and PSS estimates due to the penalisation on the curvature identifiability problem. The first additional basis is defined by more knots. Previously the number of knots was roughly 25\% of distinct data points, and we increase this to be one less than the number of distinct data points. The second additional basis includes extra columns for a periodic function for period and cohort constructed using a cyclic CRS basis \citep{Wood2017generalized} alongside the CRS basis. Both of these additional basis are not natural choices to use in practise as one would prefer to use the simple CRS spline alone.

\begin{figure}[!h]
	\centering
	\begin{minipage}{.5\linewidth}
		\centering
		\subfloat[Additional knots.]
		{\label{Fig: binomKnotsAPC_Period}
		\includegraphics[width=\textwidth]{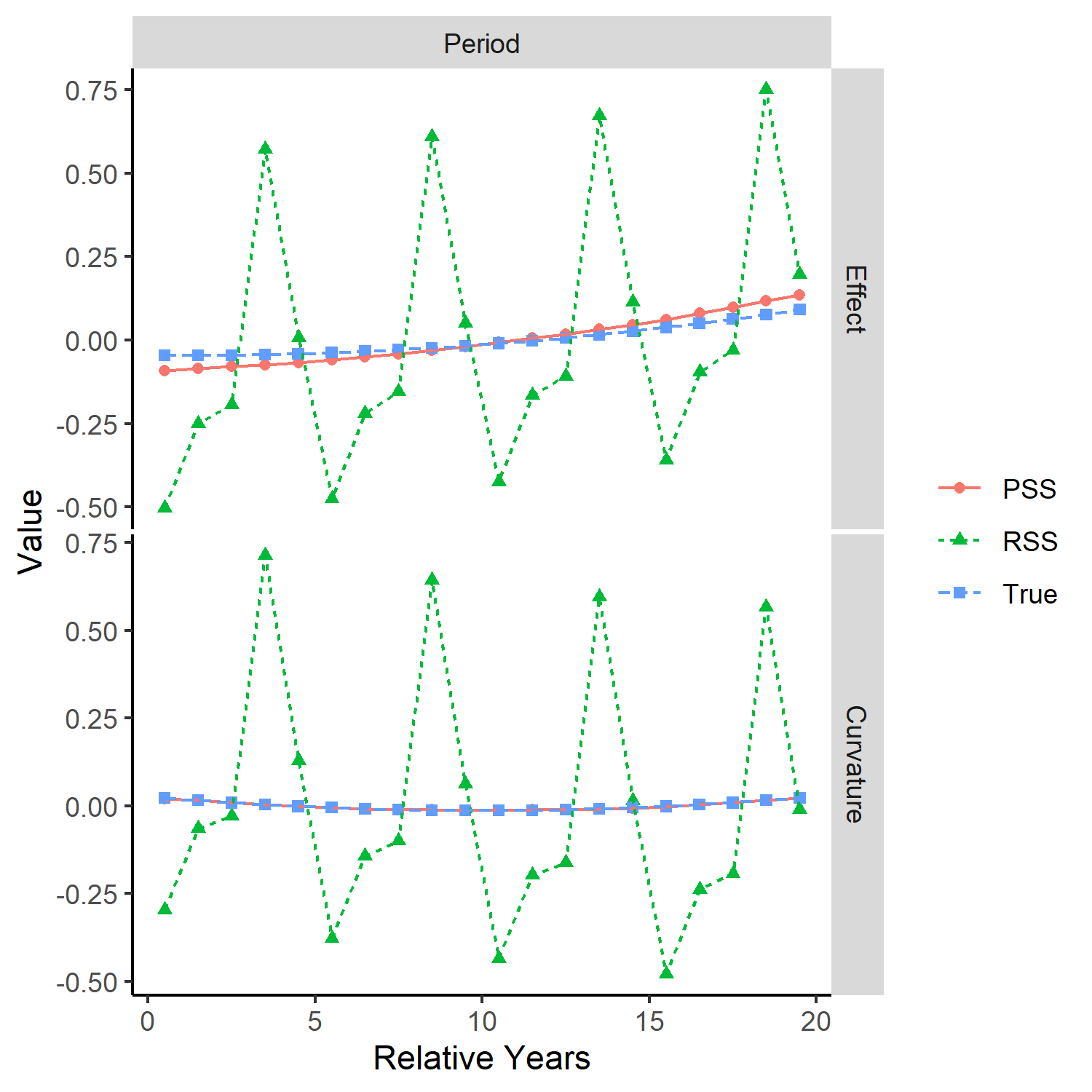}}
	\end{minipage}%
	\begin{minipage}{.5\linewidth}
		\subfloat[Inclusion of a periodic basis.]
		{\label{Fig: binomPeriodicAPC_Period}
		\includegraphics[width=\textwidth]{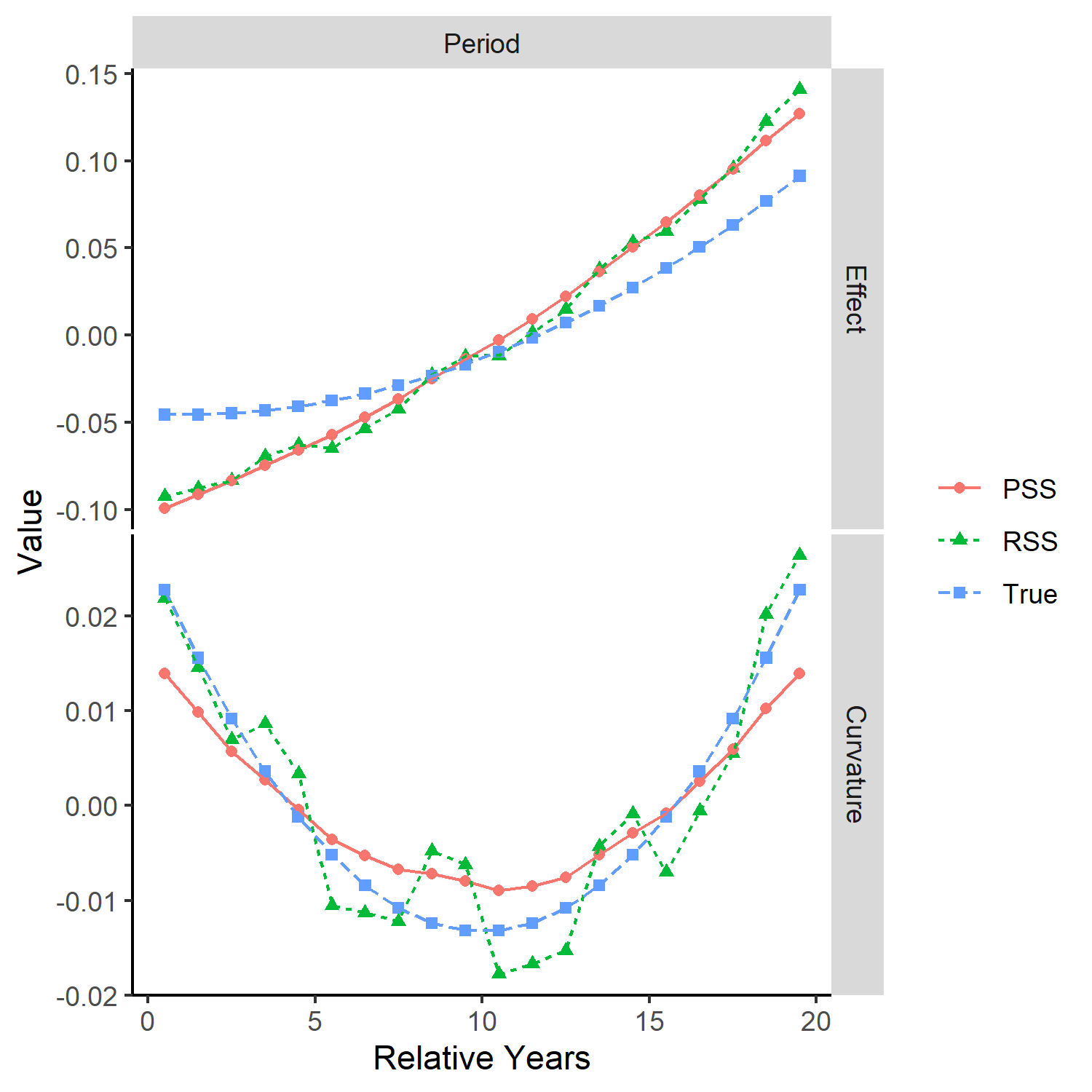}}
	\end{minipage}	
	\caption{Simulation study results for the additional bases for unequal interval, $M=5$, binomial data generated when all three temporal effects are present. Panel (a) shows the basis where the number of knots is increased to be one less than the number of distinct data points. Panel (b) shows the basis where there are additional periodic columns.}
	\label{Fig: binomAdditional}
\end{figure}

For conciseness, only the period fitted effects and curvatures against the true effects and curvatures are shown. The results from the additional simulation are in Figure \ref{Fig: binomAdditional} where \ref{Fig: binomKnotsAPC_Period} is the additional knots basis and \ref{Fig: binomPeriodicAPC_Period} is the basis with additional periodic columns. Considering the curvatures, it is clear the RSS model does not resolve the curvature identifiability problem that are more apparent when estimating with richer bases whereas the PSS model does. The penalisation for the additional knots negates all the cycling displayed by the RSS model estimates. For the additional periodic basis, the penalisation similarly removes the cycling but does over-smooth the function as a whole slightly; this is due to the non-periodic basis elements having the larger penalty also applied to them.

The results from the first simulation study show the FA model is not suitable to model data unequally aggregated in any capacity. The results from both simulation studies provide an empirical confirmation of the theoretical results. If the estimates for a given basis estimate $v_M = 0$, it gives the smoothest estimates so penalisation is not necessary and the results between the RSS and PSS models are similar. However, if the estimates for the basis estimate $v_M \neq 0$, penalisation is necessary in order to have estimates that do not suffer from the curvature identifiability problem. Furthermore, the additional simulation study results show that estimates from the RSS model are more sensitive to the choice of basis. Therefore, the user would have to perform a basis sensitivity analysis when modelling unequally aggregated APC data with an RSS model in order to be confident in their results. This is not the case for the PSS model. Results from the PSS model are not as sensitive to the choice of basis as the curvature identifiability problem is always penalised in the estimates, regardless of basis choice.

\section{Application}
\label{Section: Application}

We illustrate how the PSS model can be used in practise by applying it to UK all-cause mortality data downloaded from the Human Mortality Database (HMD). The HMD contains the raw population and (all-cause) deaths of 41 countries around the world attained from a variety of national statistic offices. The data is not shareable but is free to download after registration. \citep{HMD} Data from the HMD was chosen as it is downloadable in single-year age and period which gives the freedom to aggregate it as required.

The UK all-cause mortality data from the HMD comes in years 1922-2018 and age 0-110+. However, we take a subset of each and use years 1926-2015 and ages 0-99 to ensure equal groups when aggregating later. The HMD often receives data which is either already aggregated or contains missing values. Due to this, they fill in the missing information (using method outlined in their method protocol \citep{HMD}) which results in non-integer counts. For the purpose of modelling, the HMD values are rounded to the nearest integer and we will shall use a binomial model with a logit link function.

Figure S9 in the Supplementary Material shows a heat map of the all-cause mortality in the UK for single-year age and period in the years to be modelled. For a fixed year and in the absence of cohort effects, the apparent age effects are seen by the increase in mortality moving along the $y$-axis. For a fixed age-group and in the absence of cohort effects, the period effects appear as decrease in mortality moving along the $x$-axis. The cohort effects reflect a combination of age and period effects and appear on the bottom left to top right diagonal. An example of a notable change for each effect is: for age, the mortality changing from extremely high to low in the first five-years of life when the year is fixed at 1926; for period, the drastic reduction in mortality for age groups 0-5 in more recent periods as oppose to earlier periods; and finally, a cohort effect is the yellow to red frontier in the top right diagonally increasing due to these cohorts having a better standard of living for the entirety of their life in comparison to cohorts before. 

Three different data sets will be constructed: single-year age and period $\left(1 \times 1\right)$, five-year age and single-year periods $\left(5 \times 1\right)$ and five-year age and period $\left(5 \times 5\right)$. The $1 \times 1$ data represents the most informative data set where no aggregation occurs. The $5 \times 1$ data reflects aggregated data one might receive from a provider of health and demographic data. Finally, the $5 \times 5$ data represents data which one receives in unequal form (i.e., in $5 \times 1$) but further collapses over in order to alleviate any added identification issues that may arise in modelling. The first and last three rows of each data set can be seen in Table \ref{Tab: HMD Data}.

\begin{table}[!h]
	\caption{UK all-cause mortality data aggregated in single-year age and period, five-year age and single-year period an five-year age and period.}
	\label{Tab: HMD Data}
	\centering
	\begin{tabular}{ccccc}
	  	\hlinewd{2pt}
		\textbf{Aggregation} & \textbf{Age-Group} & \textbf{Year-Group} & \textbf{Population} & \textbf{Deaths} \\ 
		\hlinewd{1.5pt}
		\multirow{7}{*}{$1 \times 1$} & $[0,1)$ & $[1926,1927)$ & 791373 & 59661 \\ 
									 & $[0,1)$ & $[1927,1928)$ & 763981 & 56260 \\
									 & $[0,1)$ & $[1928,1929)$ & 744778 & 53281 \\
									 & $\cdots$ & $\cdots$ & $\cdots$ & $\cdots$ \\
									 & $[98,99]$ & $[2012,2013)$ & 20340 & 7751 \\
									 & $[98,99]$ & $[2013,2014)$ & 20664 & 7711 \\
									 & $[98,99]$ & $[2014,2015]$ & 40198 & 15209 \\
		\hlinewd{1.5pt}
		\multirow{7}{*}{$5 \times 1$} & $[0,5)$ & $[1926,1927)$ & 4026858 & 88081 \\ 
									 & $[0,5)$ & $[1927,1928)$ & 3888784 & 85596 \\ 
									 & $[0,5)$ & $[1928,1929)$ & 3773475 & 78393 \\ 
									 & $\cdots$ & $\cdots$ & $\cdots$ & $\cdots$ \\
									 & $[95,99]$ & $[2012,2013)$ & 90517 & 28900 \\ 
									 & $[95,99]$ & $[2013,2014)$ & 87777 & 27916 \\ 
									 & $[95,99]$ & $[2014,2015]$ & 187530 & 58471 \\
		\hlinewd{1.5pt}
		\multirow{7}{*}{$5 \times 5$} & $[0,5)$ & $[1926,1931)$ & 18960706 & 411563 \\ 
									 & $[0,5)$ & $[1931,1936)$ & 17291084 & 329432 \\ 
									 & $[0,5)$ & $[1936,1941)$ & 16790532 & 277819 \\ 
									 & $\cdots$ & $\cdots$ & $\cdots$ & $\cdots$ \\
									 & $[95,99]$ & $[2001,2006)$ & 346142 & 114044 \\ 
									 & $[95,99]$ & $[2006,2011)$ & 416010 & 131290 \\ 
									 & $[95,99]$ & $[2011,2015]$ & 457192 & 142900 \\  
		\hlinewd{2pt}
	\end{tabular}%
\end{table}

As in the simulation studies, the arbitrary choice of which slope to drop is cohort. Therefore the model equation is
\begin{equation*}
	\logit\left(\pi_{ap}\right) = \beta_0 + a\beta_{A_L} + p\beta_{P_L} + f_{A_C}\left(a\right) + f_{P_C}\left(p\right) + f_{C_C}\left(c\right).
\end{equation*}
The number of knots used for each temporal effect is 10, 10 and 20 for age, period and cohort, respectively. These are kept consistent across the models fit to all three data sets.

\begin{figure}[!h]
	\caption{UK all-cause mortality mortality fitted smooth curvatures for models fit to data aggregated in single-year age and period, five-year age and single-year period and five-year age and period. }
	\label{Fig: smooth_terms}
	\centering
	{\includegraphics[width=\textwidth]
	{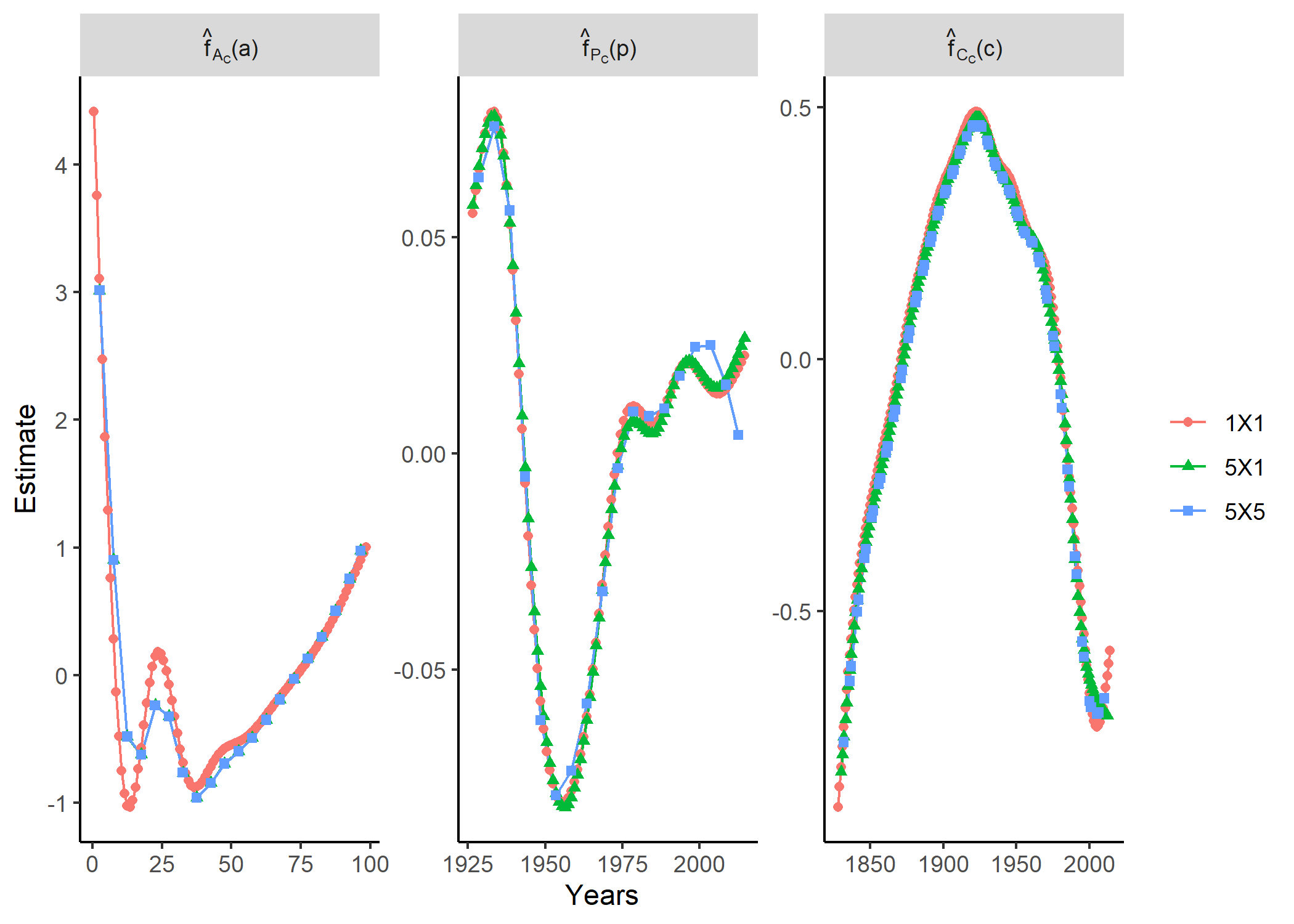}}
\end{figure}

Figure \ref{Fig: smooth_terms} shows the smooth function of the curvatures estimated from each of the data sets. These estimates are not the same as the detrended temporal estimates $\hat{h}_{\star_C}$ from the simulation studies; they are the smooth functions of temporal curvatures themselves, $\hat{f}_{\star_C}$. 

The smooth functions of curvatures represent the rate of change in a given direction. For example, the steep positively increasing half of the cohort curvature estimate reflects large improvements (large changes) in mortality in comparison to prior cohorts rather than an increase in mortality. Furthermore, the steep negatively decreasing half does not reflect a reduction in mortality rates but rather the improvements in mortality from cohort to cohort being smaller than before. In Figure S9 in the Supplementary Material, these changes can be see. The prominent diagonal frontier between the light blue and dark blue for ages 10-30 and years 1930-1960 is steeper than the frontier for 1960-present in the same age range. This means for the same ages, the apparent cohort effect reducing mortality is less pronounced. This could be from advances in living standards slowing down for the latter half of the 1900s onwards.

Given age is aggregated over five in the $5 \times 1$ and $5 \times 5$, the two sets of estimates of smooth functions are imperceptibly different to one another, hence the appearance of only two curves in the age column of Figure \ref{Fig: smooth_terms}. Both aggregated age estimates follow roughly the same trend as the un-aggregated estimates. The effect of the aggregation is clear: the more drastic changes in mortality are not captured in as much detail when aggregating. Given the slower rate of change in the cohort estimates, it is no surprise the three sets of cohort estimates are extremely similar. Each of the three functions follow a similar path, reach similar peaks and have similar start and end points.

The difference between the $5 \times 1$ and $5 \times 5$ is apparent in the estimates for the period smooth functions. At times of large change, the estimates from the $5 \times 5$ do not capture the full extent of change (e.g., the 1930s peak and 1950s trough) and have conflicting estimates (fluctuations in the 2000s). In comparison, the $5 \times 1$ model, which does not rely on collapsing to resolve added identification, follows the more informative $1 \times 1$ estimates extremely well.

Clearly, aggregating over groups loses information. The difference between the $1 \times 1$ and the other two estimates for age smooth functions show this. To then further collapse the aggregated data, from $5 \times 1$ to $5 \times 5$, loses even more information and reduces the explanatory power of the model. When data does not come in equal intervals, there is still substantial information present to give detailed representation of how mortality changes over time. This application clearly demonstrates that further collapsing to avoid complications negatively impacts the explanatory power of the model, which will impact the reason the model is being fit in the first place (evaluating interventions, policy change, analysis, etc.).

Being able to capture the larger changes in mortality for a given effect is one of the most important aspects of mortality modelling. Gaining insight into what causes these changes is helpful to understanding whether similar changes will happen again and if so, will interventions help in any way. The APC penalised smoothing spline model on unequal data produces estimates in-line with the richer data, highlighting the importance for a method that can handle data in any format.

\section{Conclusion}
\label{Section: Conclusion}

In this paper, we conducted a simulation study to investigate the use of penalised smoothing splines on the well-known `structural link' identification problem in APC models for data tabulated in equal and unequal intervals. The proposed method was compared to two different implementations of the same re-parameterisation scheme. \citep{holford1983estimation} The two other implementations are the original factor model and a smoothing spline version of the model. The penalised smoothing spline model performed in line with the current literature for data tabulated in equal intervals and outperformed the current literature for data tabulated in unequal intervals. Further benefits of an APC model that can handle unequal data were described during an application to UK all-cause mortality data from the HMD.

The results from the equal interval simulation serve as a `base-case' to demonstrate that the PSS model resolves the structural link identification appropriately by focussing on curvatures. The proposed PSS model is compared to a factor (FA) and regression smoothing spline (RSS) model. The results for the FA model from the first unequal intervals simulation highlight the curvature identifiability problem present when fitting an APC model to unequally aggregated data and demonstrate the cyclic pattern in the estimates that is the manifestation of the curvature identifiability problem. 

The results from the first unequal simulation study show little difference between the RSS and PSS models. To investigate why this is and to confirm the theoretical results, estimates of the period and cohort functions that estimate $v_M = 0$ are the smoothest, a second simulation study was conducted. In the second simulation, a richer bases was used to better estimate $v_M \neq 0$, thereby confirming if the smoothness in the first simulation was due to the model resolving the curvature identifiability problem or the choice of basis. The results of the second unequal interval simulation show the curvature identifiability problem in the RSS model but not in the PSS model. Therefore, the RSS model is sensitive to the basis functions and whether or not they reflect $v_M = 0$. Given this, care has to be taken with the RSS model as estimates are may not be consistent over a number of bases. This is not the case for the PSS model whose estimates did not display the curvature identifiability problem for the richer bases functions. Therefore, the PSS model penalised the curvature identifiability problem ensuring the estimates are identifiable.

A solution to the added complication induced by unequal intervals has been to collapse the into equal intervals. The application to UK all-cause mortality highlights how the information lost from collapsing not only yields less accurate analysis but can lead to contradictory conclusions. The results from using the unequal data reinforce how it is worthwhile to find solutions to the added complexity as there is a substantial amount of information to be gained.

Other solutions to the added complication of modelling APC model with unequal data include: using random walk (RW) priors on each of the temporal effects; \cite{riebler2009analysis} a Holford-style re-parameterisation with a factor age and continuous period term combined with a continuous full age-period interaction (cohort is a particular age-period interaction) term to model unequal APC data; \citep{heuer1997modeling} and finally, use unequal intervals for individual level data to break the structural link as no additional constraints are needed. \citep{robertson1986age, boyle1987statistical} The first alternative has theoretical parallels to our own approach as RW priors have similar penalisation characteristics \citep{Speckman2003} but does not directly address the additional identification problems present in unequal data. As with the first solution, the second solution does not directly address the additional issues as well as not being a full APC model opting for an age-period interaction over cohort. The final additional solution has since recognised by the original proponents of the approach to be an incorrect interpretation. \citep{robertson1998age}

An extension of this framework is to include forecasts. Forecasting with an APC model in a health setting is useful when updating policies and allocating resources. Consider the unidentifiable APC model where we are predicting $h$ periods into the future
\begin{equation*}
	\label{Eq: Forecasting - Unidentifiable APC Model}
	g\left(\mu_{a,p+h}\right) = f_A\left(a\right) + f_P\left(p+h\right) + f_C\left(c+h\right)
\end{equation*}
where $c = M \times \left(A - a\right) + p$. Forecasts depend on estimates, from the data, of the period and cohort functions to be projected $h$ steps ahead. When the individual temporal trends are not of interest, forecasts can be made from the above unidentifiable model. However, forecasting is more likely to be used to answer questions such as response of a given age-group over the coming years; this requires knowledge of the temporal trends. Therefore, the best practise is to perform forecasting based on invariant forecasting functions, \cite{kuang2008a} which in our proposal are the temporal curvatures.

Another extension is the inclusion of the use of non-constant unequal intervals. An example of this is the weekly period with five-year age groups from the ONS; \citep{ONS2021weekly_period} the first age group is split into 0-1 and 1-4. Another example is the Demographic Health Survey (DHS), \citep{DHS2019} a survey in low-income countries, which collects data on under-five mortality in age groups 0-1 month, 1-11 months, 12-24 months, 24-36 months, 36-48 months and 48-60 months. In the DHS, months are split like this since the first month’s deaths are extremely different to the rest of the first years; the reason for the ONS split is similar. Other APC models have been extended to incorporate covariates in space \citep{etxeberria2017spatial, chernyavskiy2019spatially} and such extensions of our work would be possible, too. The biggest challenge that will be encountered when extending our proposed APC re-parameterisation is keeping track of identifiable terms, especially when considering, for example, within covariate temporal trends. Computational challenges can arise using splines to approximate functions for large datasets (such as the DHS data) or for spatial extensions; therefore, a more appropriate way to represent the true functions as well as finding penalised estimate may need to be considered. 

\section*{Acknowledgements}

The authors would like to thank Professors Christopher Jennison and Jon Wakefield for their helpful comments on earlier drafts.

\clearpage

\bibliographystyle{unsrtnat}
\bibliography{library}

\clearpage

\end{document}